# Extremely high-resolution measurements of microwave magnetisation dynamics in magnetic thin films and nanostructures

Eugene N. Ivanov and Mikhail Kostylev
*School of Physics, University of Western Australia, Perth, WA 6009*

Abstract: In this work we discuss the use of interferometric measurement technique to study microwave magnetization dynamics on ferromagnetic nanostructures. We demonstrate that in this way one can resolve features which are impossible to resolve with broadband ferromagnetic resonance and travelling spin wave spectroscopy otherwise.

## Introduction

The concept of interferometric measurements at microwave frequencies was first suggested in the 50's, but it took more than 40 years before its high potential was fully realized. In the mid 90's, the synergy of the microwave interferometry and low-noise amplification enabled the "real-time" noise measurements with spectral resolution approaching the Standard Thermal Noise Limit (STNL) [2]. This development brought about experimental evidence of intrinsic fluctuations in the microwave components, which had been earlier considered to be "noise free". The further progress in the resolution of noise measurements went well beyond the STNL [3, 4]. This was achieved via the "power recycling" technique when studying the noise phenomena in the low-loss test samples.

The principles of microwave interferometry were also behind the breakthrough in the phase noise performance of microwave oscillators [5, 6]. Currently, the low-phase noise microwave oscillators with interferometric signal processing are a key element of the advanced Doppler radars. Apart from military applications, low-phase noise signal sources play an important role in a range of physical experiments, such as generation of entangled states between macroscopic objects and microwave photons.

The use of interferometric measurement techniques proved to be essential for understanding the origin of the excess phase noise associated with demodulation of ultra-short optical pulses produced by the mode-locked lasers [7]. Unravelling the "mystery" of the excess noise paved the way for generation of spectrally pure microwave signals from the optical sources [8, 9].

In this work we show the advantages of the interferometric measurements over conventional FMR and travelling spin wave spectroscopy when studying microwave magnetization dynamics of ferromagnetic micro- and nanostructures. We demonstrate that in this way one can resolve features which are impossible to resolve with broadband FMR and travelling spin wave spectroscopy otherwise.

## Measurement Technique

The interferometric instrument may be tuned such that it is sensitive to variation either in the amplitude or in the phase of the transmission or reflection coefficient of the device under test (DUT). This means that both characteristics can be measured with the same experimental setup and with the same accuracy. The interferometric measurements are characterized by greatly enhanced sensitivity, as compared to the conventional techniques relying on the use of a phase bridge. This is due to the ability of the interferometric systems to reconcile two seemingly contradictory requirements of having a high power incident on the DUT with a small-signal operation of its microwave readout. The enhanced sensitivity of the interferometric measurements also stems from their relative immunity to both amplitude and phase fluctuations of the output of the microwave source. This is, however, only true, if the DUT is linear and non-dispersive. For non-linear and dispersive excitations such as FMR and travelling spin waves, to minimise the fluctuations it is preferable to use a low-noise microwave generator as a source of microwave power.

An analytical expression for the smallest detectable rms fluctuations of the phase/amplitude of DUT transmission or reflection coefficient is given by

$$\delta\varphi_{\min} = \delta\alpha_{\min} \sim \sqrt{\frac{2k_B(T_0+T_i)}{P_{inc}\alpha_{DUT}}} \ ,$$

where $k_B$ is the Boltzmann constant, $T_i$ is an operational characteristic of the interferometric instrument having sense of some characteristic noise temperature, $T_o$ is the ambient temperature, $P_{inc}$ is the signal power at the input of the DUT and $\alpha_{DUT}$ is the insertion loss of the DUT. For our instrument $T_i$=50 K. This leads to almost thermal noise limited spectral resolution. Fig. 1 compares Single Sideband (SSB) phase noise floors of the interferometric and conventional (phase bridge) measurement systems. The measurements were conducted at 10 GHz with a 3 dB broadband attenuator acting as a DUT. Agilent 8257C microwave frequency synthesizer served as a pump source. As follows from the data in Fig. 1, "switching" from the phase bridge to interferometric system improves the resolution of spectral measurements by 40 dB at Fourier frequency of 100 Hz.

It should be pointed out, that the phase noise floor of the interferometric measurement system exhibits the 1/*f*-dependence at low Fourier frequencies. The influence of this technical noise can be avoided by transferring the measurements to some intermediate frequency at which spectral density of technical fluctuations falls below the fundamental thermal noise background.

Accordingly, to take FMR measurements of thin metallic films and nanostructures [10] or to carry out travelling spin wave spectroscopy of magnetic nano- and microwaveguides [11] the measurement setup is configured as shown in Fig. 2. The interferometric instrument is a single-frequency device; therefore the measurements are taken applied-field resolved. This can be repeated at a number of frequencies within the tuneability range of the instrument (approximately from 6 to 17 GHz for our instrument.) A microwave generator operational in c.w. regime is connected to the "GEN" input of the instrument. DUT is inserted between the ports "DUT". As a DUT a broadband FMR transducer in the form of a coplanar or a microstrip line with a sample on top can be used [12, 13], as well as a microwave cavity for cavity FMR or a microscopic magnetic stripe waveguide for travelling spin waves with microscopic coplanar antennas for travelling spin wave spectroscopy measurements [11].

The low-frequency output (DC OUT) of the instrument is connected to the input of a digital lock-in amplifier. The measurement data are collected from the digital output of the lock-in via GPIB. Low-frequency (220Hz and 20.2KHz respectively for our microstrip and cavity FMR setups) modulation of the applied field is utilised, and the lock-in is locked to the modulation frequency. To this end an additional modulation coil is fitted between the poles of the electromagnet. Alternatively, amplitude modulation of the output of the microwave generator may be used (which may be useful for characterisation of spin-torque nano-oscillators). However, our measurements show that the latter method is significantly less sensitive. Furthermore, the background (off-resonance) signal of non-magnetic nature is usually significant in this case, whereas it is naturally vanishing in the case of the field modulation.

Two examples of the measurements taken with the instrument are shown in Figs. 3 and 4. In both cases the instrument was tuned such that its sensitivity to variation in the amplitude of the signal from the output of DUT is maximised[1]. In these conditions the instrument is practically insensitive to the variation in the signal phase.

Fig. 3 demonstrates a broadband-FMR absorption trace taken with a 1.5mm-wide microstrip line on an array of parallel nanostripes made from permalloy ($Ni_{80}Fe_{20}$). The array was fabricated by A.O.Adeyeye's group at the National University of Singapore [10]. The macrosize of the array is 4x4 mm. The stripes are 300nm-wide, 30nm thick and 4mm long. The magnetic field is applied along the stripes. The graph demonstrates a number of higher-order standing spin waves across the stripe width. The amplitude of the outmost left-hand (negative) peak is just 20nV. From the inset one sees that no noise at all is seen in this low-applied field range. Note that these are original raw data; no graphical smoothing was used to post-process the registered trace.

The field modulation frequency and the time of stabilisation of a set magnetic field of the electromagnet determines the time for a measurement run. In our broadband FMR setup the modulation coil sits on a pole piece of the electromagnet. Because of large inductance of a coil placed on a pole piece we had to keep the frequency of field modulation low: 220Hz.
Therefore the time constant of the lock-in was set to a relatively large value of 0.3 sec. Accordingly, the time of signal accumulation by the lock-in was relatively large: 0.3s x 6 = 1.8s. This resulted in 7 minutes for completing the whole trace (200 points)[2].

---

[1] It takes 3-5 minutes to tune the instrument to a particular frequency.

[2] Note that if the coil is located well away from the pole pieces, the modulation frequency can be set much higher. Accordingly, the time for taking one measurement point can be set practically equal to the time of stabilisation of the set field of the electromagnet. This configuration is realised in our cavity FMR setup, where the coil is fixed on the wall of the cavity. In addition to the significant decrease in the measurement time, the increase in the modulation frequency further decreases the $1/f$ noise.

The second example (Fig. 4) displays the results of our measurements of higher-order modes of a microscopic stripe waveguide of travelling spin waves[3]. The $Ni_{80}Fe_{20}$ stripe has cross-section 2 micron x 100nm. The spin waves are excited and received by microscopic coplanar antennas with the total widths of 6 micron. Microscopic coplanar probes - “Picoprobes” from GGB Industries - are used to connect the antennas to the DUT ports of the instrument. Other details of the experiment can be found in Ref.[11].

The figure shows the record number of travelling spin wave modes of the waveguide taken with a fully-microwave method. In reflection from the input antenna seven modes are seen. With our simulation [14] they are identified as the lowest-order odd modes (from $n$=1 (fundamental) to $n$=13). In the signal from the output port of the spin wave device (“transmitted signal”) the $11^{th}$ mode is easily noticeable, although the distance between the antennas is quite large: 12 micron.

**Conclusion**

We have demonstrated the possibility of extremely low-noise noise measurements of microwave magnetisation dynamics of magnetic nanostructures. This was achieved by using the principles of microwave circuit interferometry.

[3] These measurements were taken by C. Chang.

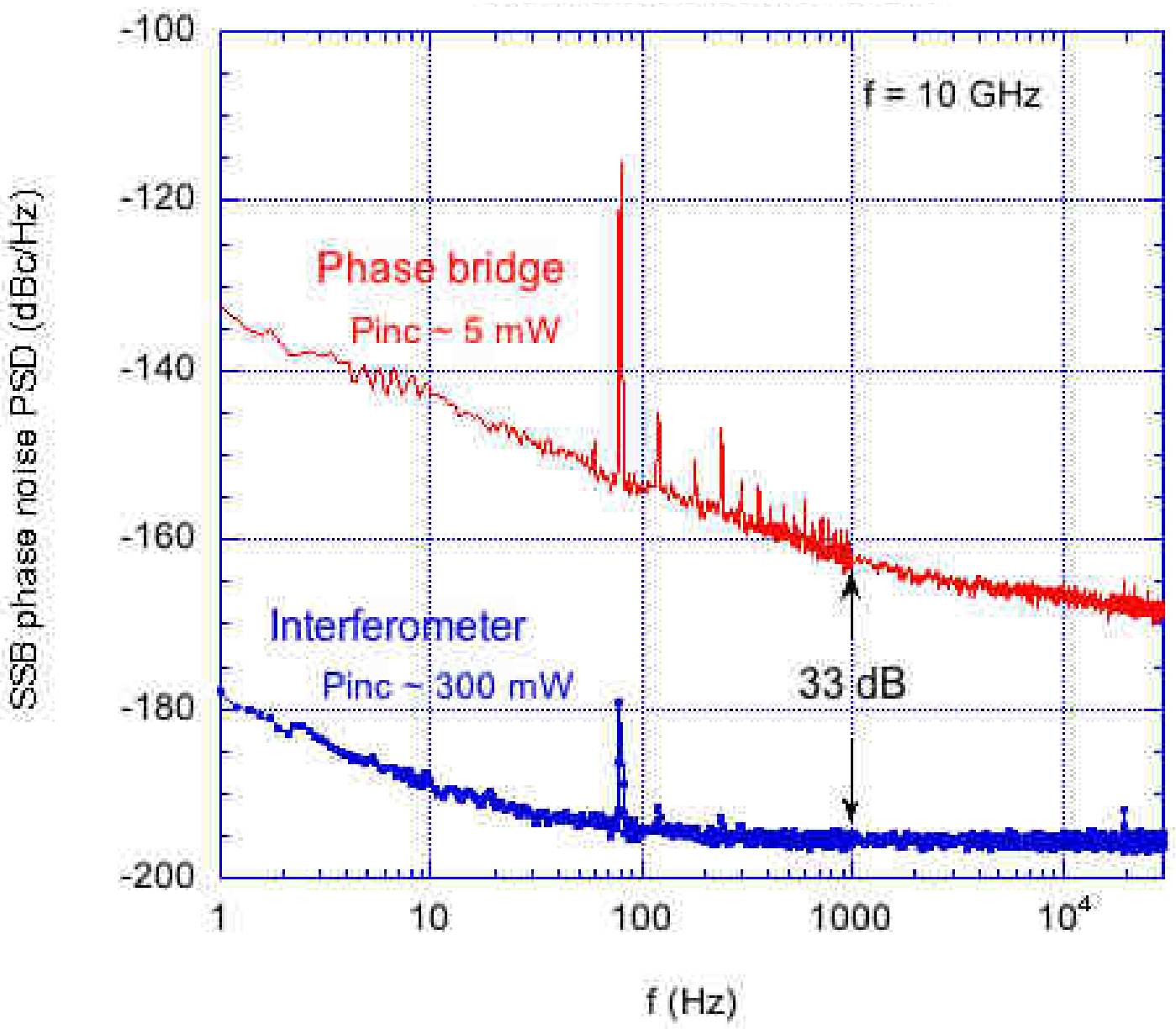

Fig. 1. SSB phase noise floors of a microwave interferometer and of an exemplary microwave phase bridge.

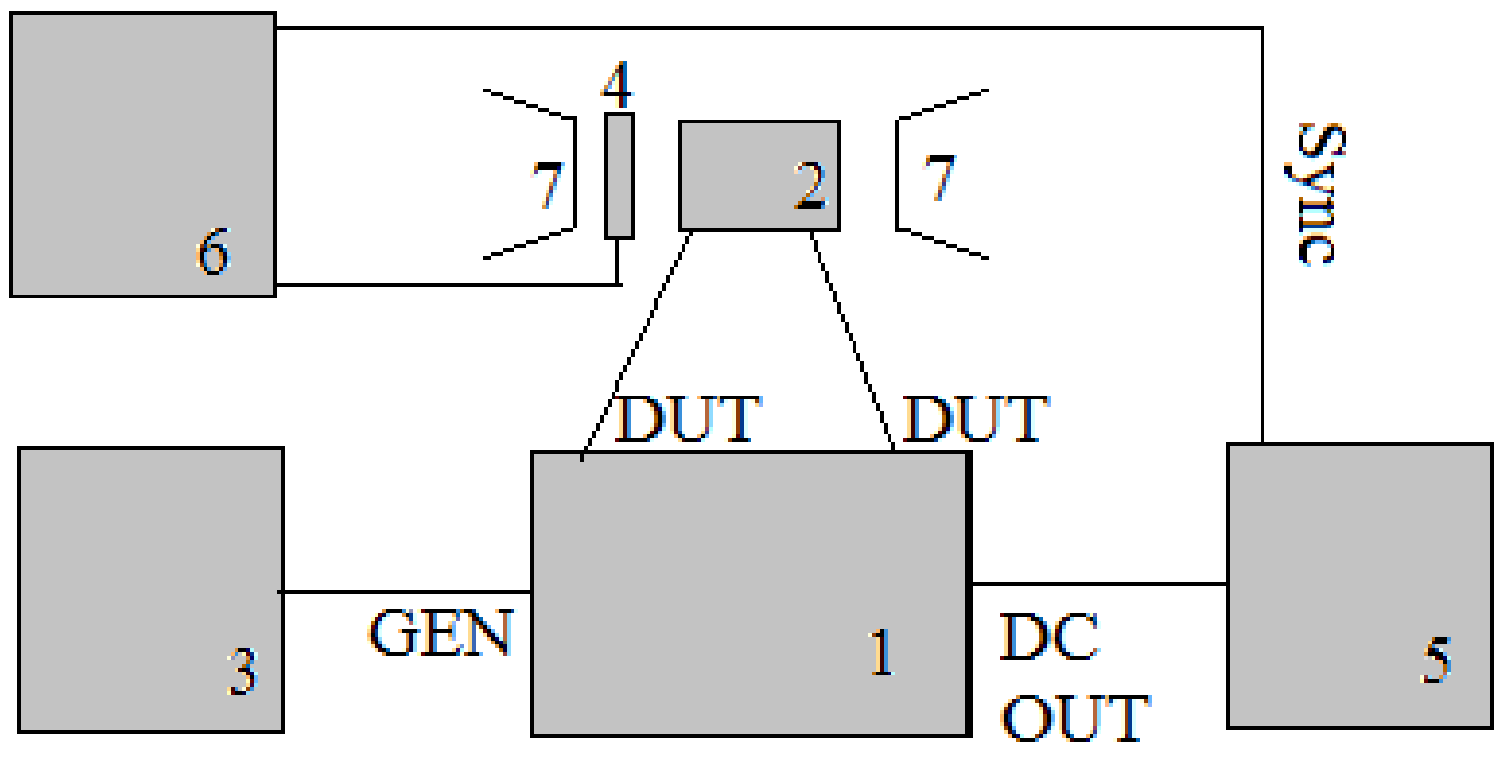

Fig. 2. Diagram of the setup for measurements of microwave magnetisation dynamics. 1. Interferometric Instrument. 2. Device under test (DUT). 3. Microwave generator. 4. Modulation coil. 5. Lock-in amplifier. 6. Function generator to provide the modulation coil with an ac current. 7. Pole pieces of the electromagnet. “Sync” denotes synchronisation signal to which the lock-in is locked.

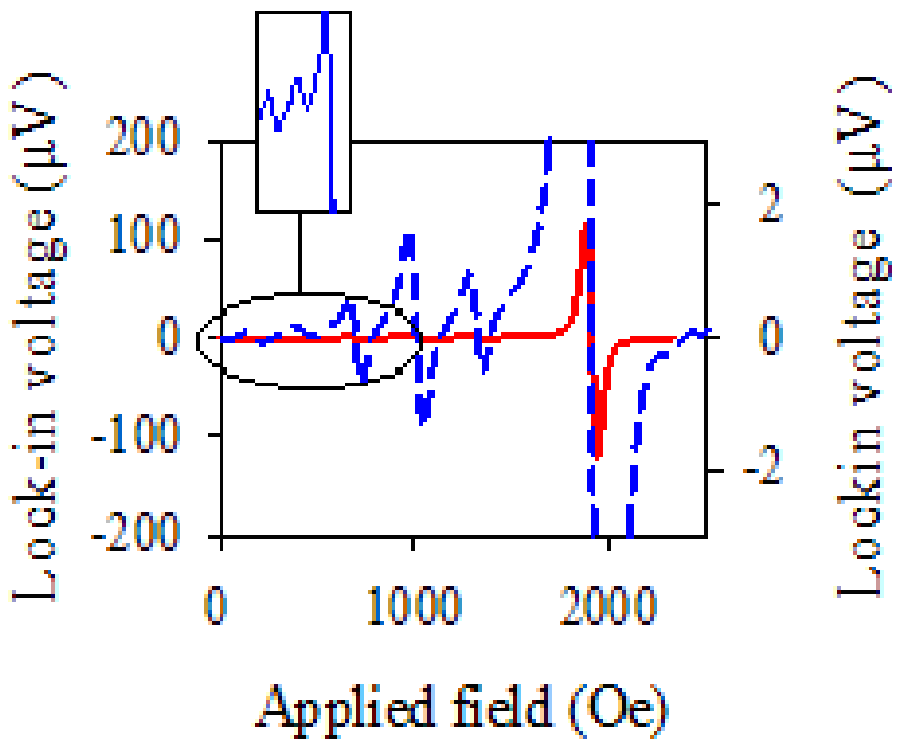

Fig. 3. Broadband ferromagnetic resonance trace for an array of permalloy nanostripes. Frequency is 15 GHz. Red solid line (left-hand axis): measurement taken with standard sensitivity of broadband FMR. Blue dashed line (right-hand axis): measurement taken with increased sensitivity of the lock-in amplifier to resolve the signal of the low-amplitude higher-order modes. Inset: zoom-in of the lower applied field range of the trace. The amplitude of the left-hand peak in the inset is 100nv (200nV peak-to-peak).

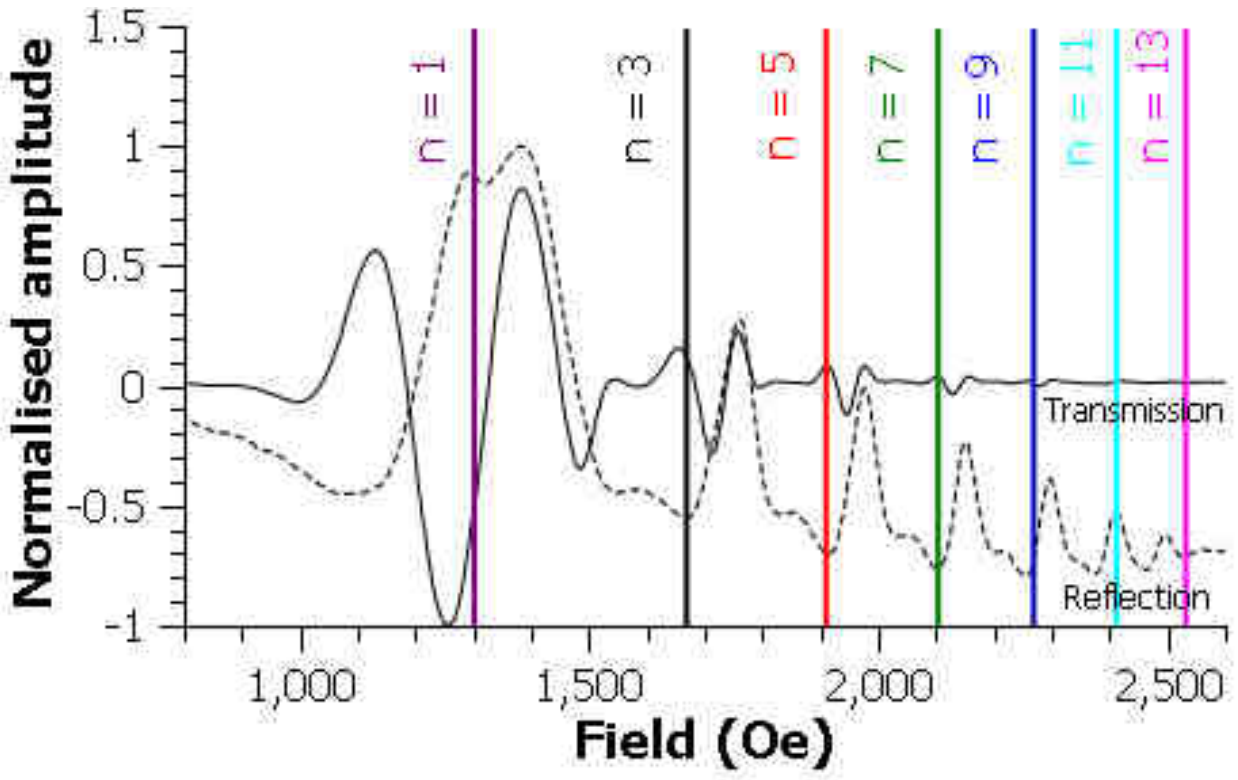

Fig. 4. Travelling spin waves on a Permaloy nanostripe with cross-section 100nm x 2 micron. Distance between the input and the output antennas is 12 micron. The field is applied in the stripe plane along the 2 micron size. Frequency is 10GHz. Dashed line: signal reflected from the input antenna. Solid line: signal from the output antenna (transmitted signal). Vertical solid lines: theoretical positions of the guided width modes for the nanostripe given here in order to identify the respective signals.